\documentclass[
reprint,
superscriptaddress,
amsmath,amssymb,
aps,
pra,
longbibliography,
]{revtex4-2}
\usepackage{float}
\usepackage{xcolor}

\usepackage{graphicx}% Include figure files
\usepackage{dcolumn}% Align table columns on decimal point
\usepackage{bm}% bold math
\usepackage[colorlinks=true, pdfstartview=FitV, linkcolor=blue, citecolor=blue, urlcolor=blue]{hyperref}
\usepackage{dsfont}

\usepackage[separate-uncertainty=true]{siunitx}

\graphicspath{{figures/}}

\newcommand{\up}{\uparrow}
\newcommand{\down}{\downarrow}

\begin{document}

\preprint{APS/123-QED}

\title{Three Strongly Coupled Kerr Parametric Oscillators Forming a Boltzmann Machine}

\author{Gabriel Margiani}
\affiliation{Laboratory for Solid State Physics, ETH Z\"{u}rich, CH-8093 Z\"urich, Switzerland.}
\author{Orjan Ameye}
\affiliation{Department of Physics, University of Konstanz, D-78457 Konstanz, Germany.}
\author{Oded Zilberberg}
\affiliation{Department of Physics, University of Konstanz, D-78457 Konstanz, Germany.}
\author{Alexander Eichler}
\affiliation{Laboratory for Solid State Physics, ETH Z\"{u}rich, CH-8093 Z\"urich, Switzerland.}
\affiliation{Quantum Center, ETH Zurich, CH-8093 Zurich, Switzerland}

\date{\today}

\begin{abstract}
  Coupled Kerr parametric oscillators (KPOs) are a promising resource for classical and quantum analog computation, for example to find the ground state of Ising Hamiltonians. Yet, the state space of strongly coupled KPO networks is very involved. As such, their phase diagram sometimes features either too few or too many states, including some that cannot be mapped to Ising spin configurations. This complexity makes it challenging to find and meet the conditions under which an analog optimization algorithm can be successful. Here, we demonstrate how to use three strongly coupled KPOs as a simulator for an Ising Hamiltonian, and estimate its ground state using a Boltzmann sampling measurement. While fully classical, our Letter is directly relevant for quantum systems operating on coherent states.
\end{abstract}

	\maketitle

%%% Introduction

The use of nonlinear oscillators for computation is a long-standing concept that has recently gained renewed interest, driven by the growing variety of available resonator platforms~\cite{Csaba_2016}. Of special interest are Kerr parametric oscillators (KPOs), bistable driven systems that can be implemented in optical, electrical, and mechanical platforms~\cite{giordmaine1965tunable, Ryvkine_2006, Mahboob_2008, Wilson_2010, Eichler_2011_NL, Gieseler_2012, Lin_2014, marandi2012all, Puri_2017, Nosan_2019, Frimmer_2019, Grimm_2019, wang_2019, Puri_2019_PRX, Miller_2019_phase, yamaji_2022,Yamaji_2023,Frattini_2024,Hoshi2025}. In a KPO, a modulation of the resonator potential energy, which we refer to as parametric pumping, gives rise to two possible ``phase states'' that have identical amplitudes and opposite phases~\cite{Dykman_1993,lifshitz2008nonlinear,Eichler_Zilberberg_book}. Early on, these states were proposed and used as a physical basis for digital computing~\cite{Goto_1959,Neumann_1959,Sterzer_1959}.

\begin{figure}[t]
    \includegraphics[width=\columnwidth]{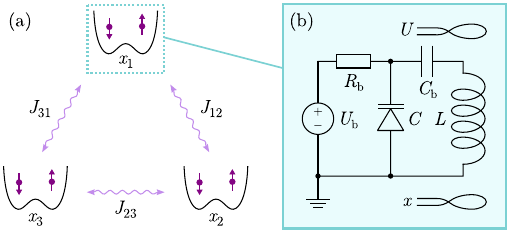}
    \caption{Ising network built from three coupled KPOs. (a)~The parametric phase states are represented as double-well quasipotentials whose minima are our artificial Ising spins. (b)~Single KPO realized as a RLC circuit. We use a varactor diode as nonlinear capacitance $C$ and a coil as inductance $L$. The capacitor $C_\mathrm{b}$ and resistor $R_\mathrm{b}$ decouple coil and bias source. We use the voltage $U = U_\mathrm{F}\cos(\omega t) + U_\mathrm{p}\cos(2\omega t)$ to drive the resonator, while we measure the voltage signal $x$.}
    \label{fig:fig1}
\end{figure}

The KPO is currently a major research focus due to the analogy between its two phase states and the ``up'' and ``down'' polarization states of an Ising spin. In particular, it was proposed that networks of KPOs [see Fig.~\ref{fig:fig1}(a)] can be used to find the ground state of Ising Hamiltonians, that is, the energetically preferred configuration of a spin network~\cite{Ising_1925}. Such resonator-based Ising solvers~\cite{Mahboob_2016, Goto_2016, Dykman_2018, Puri_2019_PRX, Bello_2019, Okawachi_2020,yamamoto2020coherent,Han_2024,Boehm2025,Tosca2024arxiv} are of high interest because the corresponding calculations are hard to tackle with conventional computers~\cite{mohseni2022ising}. At the same time, they map to many key optimization problems, such as the traveling salesman problem~\cite{Lucas_2014}, the MAX-CUT problem~\cite{Inagaki_2016_Science, Goto_2019}, and the number partitioning problem~\cite{Nigg_2017}. Interestingly, in the quantum regime, a KPO network can function as a quantum annealer, potentially outperforming its classical counterpart in finding optimal solutions~\cite{kadowaki_1998,Goto_2016}. To maintain quantum coherence during the adiabatic annealing procedure, the coupling between the $N$ nodes must be stronger than the decay rate. Strong coupling, in turn, leads to a complicated phase diagram that does not necessarily map to an Ising Hamiltonian~\cite{heugel2019classical,Heugel_2022,Alvarez_2024}. So far, there exists no experimental confirmation that strongly coupled KPO resonators can function as Ising simulators. 

In this Letter, we demonstrate Ising simulations on a network of three strongly coupled KPOs. To do so, we explore the phase diagram of our network, and identify where in parameter space it possesses the correct number of solutions to manifest an Ising Hamiltonian.
In that region, we perform Boltzmann sampling~\cite{Goto_2018,Margiani_2023} and extract the total dwell times the system spends in each state. This procedure allows us to find the correct ground state of the corresponding Ising Hamiltonian, both without and with an applied external bias field. Finally, we highlight the crucial role of the nonlinearity and its relation to the rotating-frame Hamiltonian. Our Letter thus establishes that Ising simulation is viable in the strong coupling limit and provides guidelines for optimized experimental setups.

%%% The Setup

Our experimental setup consists of three inductively coupled nonlinear resistor-inductor-capacitor (RLC) resonators, cf. Fig.~\ref{fig:fig1}(b). Each resonator uses a varactor diode as a nonlinear element, which allows tuning the (angular) resonance frequency $\omega_0$ via a reverse dc voltage $U_\mathrm{b}$. External forcing and parametric pumping is implemented via a wire loop close to the main inductor of the circuit. Applying an ac voltage tone $U(\omega)$ to the drive loop induces a current in the resonator, which acts as a near-resonant forcing term with amplitude $U_\mathrm{F}$ when $\omega\approx\omega_0$. At the same time, the second-order nonlinearity $\beta_2$ introduced by the varactor diode (analogous to an optical $\chi^{(2)}$ nonlinearity) enables three-wave mixing and parametric pumping with a modulation depth $\lambda$~\cite{Nosan_2019,Eichler_Zilberberg_book}. When a voltage with amplitude $U_\mathrm{p}$ is applied close to $2\omega \approx 2\omega_0$, parametric oscillation can arise if $\lambda$ exceeds a threshold $\lambda_{\mathrm{th}}$. These oscillations induce a voltage $x$ in a second readout loop. The voltage is measured using an Intermodulation Products multifrequency lock-in amplifier.

We model our system by the following set of coupled differential equations, each describing a nonlinear parametric oscillator:
\begin{multline}
	\ddot{x}_i + \omega_0^2\left[1-\lambda\cos\left(2 \omega t\right)\right]x_i + 
 \beta x_i^3 + \Gamma \dot{x}_i
 - \sum_{j\neq i}  J_{ij} x_j = F_i\,\label{eq:EOM}
\end{multline}
where $x_i$ is the voltage measured from each of the resonators $i \in \{1, 2, 3\}$ with an effective Duffing nonlinearity constant $\beta = -10\beta_2^2/(9\omega_0^2)$~\cite{lifshitz2008nonlinear,Eichler_Zilberberg_book}, a damping rate $\Gamma$, and an external forcing term $F_i=F\cos(\omega t)$ with $F\propto U_\mathrm{F} $~\cite{Nosan_2019,Heugel_2022}. The parametric pump modulates the resonator's potential with a modulation depth $\lambda = \frac{2 U_\mathrm{p}}{Q U_\mathrm{th}}$, where $U_\mathrm{p}$ is the amplitude of the parametric drive signal, $U_\mathrm{th}$ the parametric threshold voltage, and $Q$ the quality factor of the resonator. Finally, each resonator couples to the two others via symmetric coupling constants $J_{ij} = J_{ji}$.

We individually characterize each of the resonators while keeping the others far detuned (by setting their bias to $U_{\mathrm{b}i} = 0$). This allows us to extract their bare parameters independently of coupling effects, and to calibrate the values of $U_{\mathrm{b}i}$ required to tune all resonators to the same frequency $\omega_0$ (and roughly the same $\Gamma_i = \Gamma$). Using $U_{\mathrm{b}1} = \SI{5.034}{\volt}$, $U_{\mathrm{b}2} = \SI{5.0239}{\volt}$, and $U_{\mathrm{b}3} = \SI{5.0}{\volt}$, we find the following average parameters, with errors indicating device-to-device differences: $\omega_0/2\pi = \SI{3224100(50)}{\hertz}$, $U_\mathrm{th} = \SI{48.5(1.5)}{\milli\volt}$, $\beta = \SI{-35(1)e15}{\hertz\squared\per\volt\squared}$, $\Gamma = \SI{49.45(0.15)}{\kilo\hertz}$, and $Q = \num{409.7(1.2)}$. For most of this Letter, we additionally assume identical coupling constants $J = J_{ij} = -\SI{1107(4)e9}{\hertz\squared}$, which we extracted from the frequency splitting observed in the parametric response of the coupled resonators.

\begin{figure}[t]
    \includegraphics[width=\columnwidth]{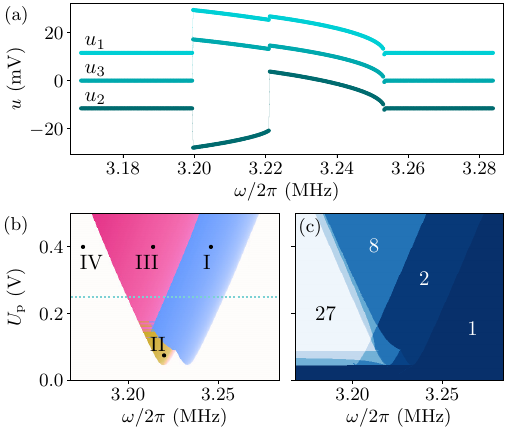}

    \caption{Frequency sweeps. (a)~Response of the three resonators to a frequency sweep with $U_\mathrm{p} = \SI{250}{\milli\volt}$ and $U_\mathrm{F} = 0$. Only the $u$ (in-phase) quadrature is shown. The signals of resonator 1 and 2 have been shifted by an offset $\pm\delta_\mathrm{u}$ for better visibility. (b)~Phase diagram of the coupled system, measured as frequency up-sweeps at different $U_\mathrm{p}$. Shade qualitatively shows the response amplitude from low (bright) to high (dark). Colors indicate the configuration. Blue (around I): all three resonators move in phase with roughly equal amplitude. Yellow (around II): only two resonators have a non-zero amplitude and oscillate with roughly opposite phases. Magenta (around III): two resonators have the same phase, while the third one oscillates with opposite phase. White (remaining space): all resonators have zero amplitude. Labeled points mark the positions used in Fig.~\ref{fig:fig3}. (c)~Number of stable stationary solutions of Eqs.~\eqref{eq:EOM} calculated by Harmonic Balance~\cite{kovsata2022harmonicbalance}, encoded in the brightness contrast.}
    \label{fig:fig2}
\end{figure}

We start our experimental investigation of the KPO network with $F=0$. We tune all KPOs into resonance and pump them with the same parametric drive $U_\mathrm{p}\cos(2\omega t)$, see Fig.~\ref{fig:fig2}(a). We plot the $u$ quadrature of the oscillation, defined by $x_i=u_i\cos(\omega t) + v_i\sin(\omega t)$, where $u$ and $v$ are the quadratures measured by our lock-in amplifiers. The $v$ quadrature yields analogous results. Sweeping $\omega$ from low to high frequencies across $\omega_0$, we observe that the amplitudes of all three resonators jump from zero to a large value at $\omega/2\pi \approx \SI{3.2}{\mega\hertz}$. Close to $\omega/2\pi \approx \SI{3.22}{\mega\hertz}$, we observe a second jump, followed by a gradual decrease of all amplitudes towards zero.
For each resonator, we can identify positive (negative) $u_i$ with the up-state $\up_i$ (down-state $\down_i$) of an Ising spin, while the zero-state $0_i$ with $u_i \approx 0$ is outside of the Ising spin model. Accordingly, we measure in Fig.~\ref{fig:fig2}(a) the states $(0_1\,0_2\,0_3)$, $(\up_1\,\down_2\,\up_3)$, $(\up_1\,\up_2\,\up_3)$, and $(0_1\,0_2\,0_3)$ along the $\omega$ sweep. In the following, we drop the subscripts from the state labels.

Measuring frequency up-sweeps for various $U_\mathrm{p}$ allows us to sample different states in a phase diagram, analogous to the Arnold tongue of a single KPO; see Fig.~\ref{fig:fig2}(b)~\cite{Heugel_2022}. Here, colors indicate the resonator phase combinations corresponding to different state configurations. The blue region indicates $(\up\up\up)$ or $(\down\down\down)$. In the magenta region, one ``spin'' is opposite to the other two, e.g., $(\up\up\down)$. In the yellow region, we find $(0\!\up\down)$ and permutations thereof, which are not Ising states. The measurement protocol tracks the system as it rings down into a single state configuration at each position in the phase diagram in a deterministic fashion.

\begin{figure}[t]
    \includegraphics[width=\columnwidth]{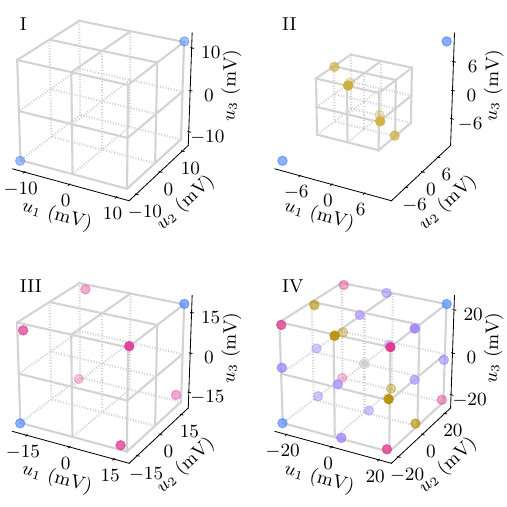}
    \caption{Oscillation states measured at different positions in Fig.~\ref{fig:fig2}(b). An external force at the measurement frequency $\omega$ is applied to each resonator before the parametric pump is switched on. By varying the force phases, different initial conditions are generated, triggering the system to choose different stationary oscillation solutions. The external forces are switched off when the parametric pump is switched on.
    }
    \label{fig:fig3}
\end{figure}

Crucially, a network with $N=3$ KPOs can have up to  $3^N = 27$ solutions; see Fig.~\ref{fig:fig2}(c)~\cite{borovik2024khovanskii, ameye2025parametricinstabilitylandscapecoupled}. To identify these additional states, we apply an external force to preset the resonators with different phases before activating the parametric pump, and we systematically measure the final states across various initial conditions; see Fig.~\ref{fig:fig3}. With this procedure, we find two states at position I in Fig.~\ref{fig:fig2}(b), eight states at positions II and III, and 27 states at position IV, in agreement with the model prediction; cf.~Fig.~\ref{fig:fig2}(c). Both positions II and III feature the correct number of states to represent all solutions of an Ising network ($2^N = 8$). However, most of the states at II contain zero-amplitude oscillations, which do not correspond to any Ising state. Only position III has the correct number of nonzero states to represent the full Ising state space.

\begin{figure*}[t]
    \includegraphics[width=\textwidth]{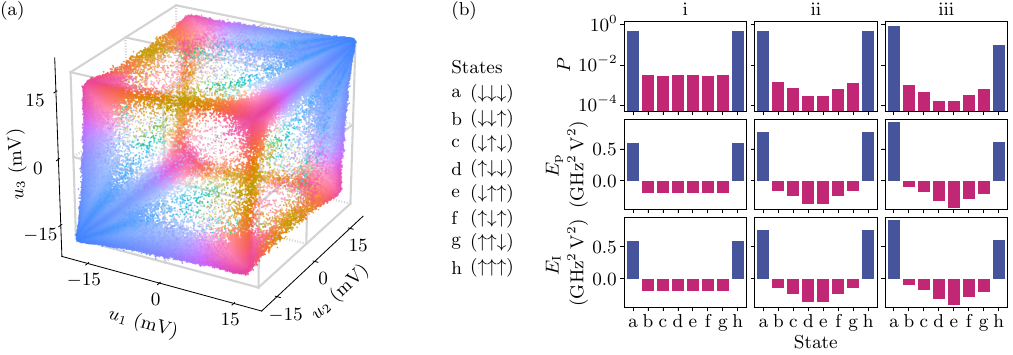}
    \caption{Stochastic sampling of the network. The resonator system is driven parametrically at point III in Fig.~\ref{fig:fig2}(b). Additionally, a white noise signal with a standard deviation of $\SI{137.88}{\milli\volt}$ low-pass filtered at \SI{5}{\mega\hertz} is applied to the drive of each resonator, triggering switches between different phase states. The resonator's response is measured for $t_\mathrm{m} = \SI{400}{\second}$
    %, acquiring $\mathit{Sa} = 8032129$ samples
    with a sampling rate of $\mathit{df} = \SI{20.081}{\kilo\hertz}$. (a)~3D representation of the $u$ quadratures of a full dataset. Color encodes the configuration of the state as in Fig.~\ref{fig:fig3}. Corners correspond to stable states, and jumps between them occur predominantly along the edges of the cube. The difference between the outermost blue and pink points appears exaggerated (relative to Fig.~\ref{fig:fig3} III) due to the presence of noise and the difference in state occupation probability $P$. (b)~Ising ground state prediction algorithm. Upper row shows the measured occupation probability; cf. data in (a) and \cite{supmat} for details. The left column (i) is for identical coupling, the middle column (ii) is for $J_{12} < J_{31} < J_{23}$, and the right column (iii) has the same coupling as in the middle with an additional external force of $U_\textrm{F} = \SI{5}{\milli\volt}$ to break the symmetry between the phase states of each KPO. The middle and lower rows show the quasienergies $E_\textrm{p}$ of the KPO network in the rotating frame [cf. Eq.~\eqref{eq:quasi_energy}] and the eigenenergies $E_\textrm{I}$ for analogous configurations in a three-spin Ising Hamiltonian [cf. Eq.~\eqref{eq:Ising}].}
    \label{fig:fig4}
\end{figure*}

Having established that position III in Fig.~\ref{fig:fig2}(b) features the correct states to serve as an Ising solver, we demonstrate a protocol to identify the corresponding Ising ground state from the KPO network. To remove the dependence on the initial condition, we implement Boltzmann sampling~\cite{Goto_2018,Margiani_2023}: applying white force noise results in activated jumps between the possible states of the KPO network~\cite{Dykman_1993,Chan_2008}, which we measure in real time and plot in Fig.~\ref{fig:fig4}(a). Switching events are rare compared with the duration of a switch, hence the system predominantly switches between states by inverting the phase state of a single resonator at a time. Such switching events, for example from $(\up\up\up)$ to $(\up\up\down)$, take place along the edges of the cube in Fig.~\ref{fig:fig4}(a). Double switches, for example from $(\up\up\up)$ to $(\up\down\down)$, usually involve two sequential single switches instead of a direct diagonal line across a face of the cube.

Boltzmann sampling of a KPO network was proposed as a way to find the ground state of an Ising Hamiltonian~\cite{Goto_2018}. To see whether this idea holds in our network, we extract from Fig.~\ref{fig:fig4}(a) the occupation probability $P(\sigma_1,\sigma_2,\sigma_3) = \frac{\tau}{t_\mathrm{m}}$ of each state, with $\tau$ the total time spent in the state and $t_\mathrm{m}$ the measurement duration. See top row in Fig.~\ref{fig:fig4}(b). We perform this analysis for three different cases, with (i)~identical all-to-all coupling $J$, (ii)~nonequal negative coupling coefficients $J_{12} < J_{31} < J_{23}$, and (iii)~nonequal coefficients with an additional external force $F>0$ that breaks the symmetry between the phase states~\cite{Ryvkine_2006,Leuch_2016,Alvarez_2024,boness2024resonant}. The third example introduces an analogy to a bias field $B$ in the Ising Hamiltonian. We find systematic differences between the values of $P$ in all three cases.

Trying to understand the switching dynamics in our KPO network, we derive its effective quasi-Hamiltonian; see \cite{supmat} for details. In the absence of damping, this quasi-Hamiltonian is the function that governs the deterministic motion of our system in the frame rotating at $\omega$~\cite{Dykman_1998,Seibold_2025,Goto_2016,Dumont_2024,Eichler_Zilberberg_book}. Without loss of generality, we align this rotating frame such that the phase states of each resonator $i$ appear at $\phi_i \in \{0, \pi\}$, and we indicate with $\sigma_i = \cos{(\phi_i)} \in \{\pm 1\}$ and $A_i$ the corresponding phases and amplitudes of the measured voltage $x_i$, respectively. In this way, we arrive at a form $H_\mathrm{eff} = H_0 +H_\mathrm{J}$, where the decoupled part is equal to
\begin{align}\label{eq:decoupled}
  H_0 =\sum_i \frac{3}{32} \beta A_i^4 - \frac{1}{8} \left( 2 \omega^2 - 2 \omega_0^2 + \lambda\omega_0^2 \right) A_i^2 - \frac{F A_i}{2}\sigma_i,
\end{align}
while the coupling Hamiltonian simplifies to
\begin{equation}
  H_\mathrm{J} = -\sum_{j \neq i} \frac{J_{ij}}{2} A_i A_j \sigma_i \sigma_j.
\end{equation}
Assuming identical $A_i=A$ for simplicity, the quasienergy differences between the stationary states follow from
\begin{align}\label{eq:quasi_energy}
E_\mathrm{p}=-\frac{A^2}{2}\sum_{j>i}J_{ij}\sigma_i\sigma_j-\frac{FA}{2}\sum_i \sigma_i.
\end{align}
These quasienergies are shown in the middle row of Fig.~\ref{fig:fig4}(b).

In the lowest row of Fig.~\ref{fig:fig4}(b), we show the eigenenergies $E_\textrm{I}$ of the Ising Hamiltonian,
\begin{align}\label{eq:Ising}
E_\textrm{I}=-\sum_{j>i} W_{ij}s_i s_j -B_\mathrm{I}\sum_{i} \sigma_i,
\end{align}
where $W_{ij}$ are coupling energies and $s_{i}\in\{\pm 1\}$ correspond to Ising spin states $\up$ and $\down$. From a comparison between Eqs.~\eqref{eq:Ising} and \eqref{eq:quasi_energy}, we can determine the values of $W_{ij}$ and $B_\mathrm{I}$ that our KPO network maps to.

When analyzing all three rows, we find a strong correlation between $E_\textrm{p}$ and $E_\textrm{I}$. In particular, the quasi-Hamiltonian model correctly predicts the ground state of the Ising Hamiltonian in all cases. Furthermore, we can use $P$ to experimentally predict the ground state, confirming the validity of the Boltzmann sampling method. However, we systematically measure \textit{lower} values of $P$ for lower $E_\textrm{p}$. This can appear paradoxical at first: after all, in an equilibrium system one would always expect \textit{higher} $P$ for lower eigenenergies.

\begin{figure}[t]
    \includegraphics[width=\columnwidth]{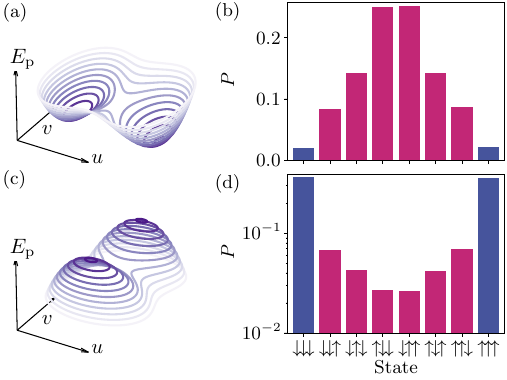}
    \caption{Inverting the nonlinearity. (a)~Schematic of the quasipotential for a single parametric oscillator with positive $\beta$. (b)~Numerical analysis of occupation probabilities $P$ for three resonators with $\omega_0 = 1$, $\lambda = 0.6$, $Q = 10$ and $\beta = 1$ and coupling $J_{12} = -1.5$, $J_{23} = -0.5$ and $J_{31} = -1$. Combined results for 30 numerical simulations of $t_\textrm{m} = \SI{27.7}{\hour}$ each, with a sampling rate of $df = \SI{5}{Hz}$. (c)~Same as (a) for $\beta = -1$, and (d) same as (c) for negative $\beta$.}
    \label{fig:fig5}
\end{figure}

In a previous derivation of Boltzmann sampling with KPO networks, the case $\beta>0$ was explicitly considered~\cite{Goto_2018}. In this case, all individual KPO phase states are minima in $E_\mathrm{p}$ in terms of $u$ and $v$; see Fig.~\ref{fig:fig5}(a) and Eq.~(S7) in \cite{supmat}. Here, the state with the lowest $E_\mathrm{p}$ is the one with the highest $P$, in perfect analogy to an equilibrium system. We show the results of a numerical simulation of this situation in Fig.~\ref{fig:fig5}(b). By contrast, our experimental system, similar to KPOs in Josephson superconducting circuits~\cite{Grimm_2019}, has $\beta<0$ and all KPO phase states appear as maxima in $E_\mathrm{p}$; see Fig.~\ref{fig:fig5}(c). As a consequence of inverting the sign of $\beta$, the numerical simulation now predicts higher $P$ for higher $E_\mathrm{p}$, see Fig.~\ref{fig:fig5}(d).

To reach a deeper understanding of the role of $\beta$, we recall that the quasi-Hamiltonian in a rotating frame has different properties than an equilibrium Hamiltonian. Namely, both minima and maxima of $E_\mathrm{p}$ are stable states, and it is not clear which one should have a higher $P$.
To answer this question, we form a new effective Hamiltonian $\tilde{H}_\mathrm{eff}=H_\mathrm{eff}\times \mathrm{sign}(\beta)$ whose quartic term ($A_i^4$) is always positive (but may have a detuning-dependent offset and shifted extrema positions in phase space; see Figs.~\ref{fig:fig5}(a) and \ref{fig:fig5}(c) and Sec.~S6 in \cite{supmat}). This transformation makes every solution a minimum, allowing us to always apply the intuition from equilibrium physics, where lower eigenenergies lead to higher $P$. However, for $\beta<0$ we now have inverted the sign of $\tilde{H}_\mathrm{J} = H_\mathrm{J}\times \mathrm{sign}(\beta)$, thereby changing the ordering of the eigenenergies. In conclusion, inverting the sign of $\beta$ not only interchanges maxima and minima in the quasienergy landscape, but also effectively maps the solved Ising problem to one with couplings $J \rightarrow -J$. This confirms that, for $\beta<0$, the state with the highest eigenenergy has the highest $P$.

In summary, we use three strongly coupled KPOs for Boltzmann sampling to predict the ground state of an Ising Hamiltonian. In addition, we shed light on the relationship between the signs of the nonlinearity, the coupling coefficients, and the measured state probabilities. This will allow us to optimize the Boltzmann sampling statistics in future experiments. For example, it is clear from Fig.~\ref{fig:fig5} that the case $\beta>0$ and $J_{ij}<0$ will cause the system to spend most of the time in states near the ground state of the corresponding Ising model. Compared with $\beta<0$ and $J_{ij}<0$, this will lead to better Boltzmann sampling statistics, allowing for more efficient and certain identification of the system's ground state. For $J_{ij}>0$, the situation is reversed, and $\beta<0$ yields better sampling near the ground states. Alternatively, we can implement the inverse problem by inverting the signs of all $J_{ij}$, which has the same effect as inverting the sign of $\beta$ (see \cite{supmat} for a schematic list of possible cases). This insight allows us to optimize the Boltzmann sampling method according to the experimental constraints, and to analyze the results correctly.

\section*{Acknowledgements}
We thank Letizia Catalini for helpful discussions and Shobhna Misra for help in building the setup.
We further acknowledge funding from the Deutsche Forschungsgemeinschaft (DFG)
via Project No.~449653034 and through SFB1432, as well as the Swiss National Science Foundation (SNSF) through the Sinergia Grant No.~CRSII5\_206008/1.

\nocite{lifshitz2008nonlinear,Eichler_Zilberberg_book,margiani2021fluctuating,Goto_2018,burshteinHamiltonian1962,Seibold_2025,holmesSecond1981,ameye2025parametricinstabilitylandscapecoupled}

\bibliography{aipsamp}% Produces the bibliography via BibTeX.

\end{document}

\title{\textbf{Supplemental Material for:}\\Three strongly coupled Kerr parametric oscillators forming a Boltzmann machine}

\author{Gabriel Margiani}
% \thanks{Authors contributed equally}
\affiliation{Laboratory for Solid State Physics, ETH Z\"{u}rich, CH-8093 Z\"urich, Switzerland.}
\author{Orjan Ameye}
% \thanks{Authors contributed equally}
\affiliation{Department of Physics, University of Konstanz, D-78457 Konstanz, Germany.}
\author{Oded Zilberberg}
\affiliation{Department of Physics, University of Konstanz, D-78457 Konstanz, Germany.}
\author{Alexander Eichler}
\affiliation{Laboratory for Solid State Physics, ETH Z\"{u}rich, CH-8093 Z\"urich, Switzerland.}
\affiliation{Quantum Center, ETH Zurich, CH-8093 Zurich, Switzerland}

\date{\today}

\maketitle

\section{Experimental Setup}

% Cap 1nF, 46MOhm, 10 MOhm in drive, 
% 43-18 pF 2-8V ~27pF @ 5V -> ~45uH Coil

Our experimental setup consists of three RLC resonators coupled inductively to each other. The circuit of a single resonator is shown in Fig.~1(b) of the main text. The core resonator is formed by a large coil with inductivity $L \approx \SI{45}{\micro\henry}$ in parallel with two varactor diodes (BB914) acting as voltage dependent capacitor. The diodes are held in reverse voltage conditions across the whole oscillation cycle of the resonator by a dc bias voltage $U_\mathrm{b}$. Changing this voltage also changes the average capacitance of the varactor, thereby tuning the resonance frequency of the system ($C \approx \SI{54}{\pico\farad}$ at $U_\mathrm{b} = \SI{5}{\volt}$). The bias is decoupled from the resonator using a dc-block capacitance $C_\mathrm{b} = \SI{1}{\nano\farad} \gg C$ and a resistor $R_\mathrm{b} = \SI{46}{\mega\ohm}$ to keep the current flow negligible.

Each individual resonator can be inductively driven by applying a varying voltage signal to a wire loop close to the main coil. The wire loop is terminated by a \SI{10}{\mega\ohm} resistor. Similarly, measurements are performed using a pickup loop that is inductively coupled to the resonator. Varying the distance and shape of drive and pickup loop, the effective drive strength and measured amplitude can be tuned to compensate for differences from fabrication. The damping of the resonator can be adjusted by varying the distance of a metal nut to each coil. Eddy currents in the metal nuts dissipate energy, thereby adding to the damping. We use an \emph{Intermodulation Products MLA} (multi-frequency lock-in amplifier) to generate the drive signals and bias voltages, as well as to measure the responses. A single drive signal is distributed to the resonators via a set of power splitters. For stochastic driving, each of the three parametric drive signals is added to a unique white noise signal from a separate voltage noise source. The noise signals are low-pass filtered with a 3 db frequency of \SI{5}{\mega\hertz} to reduce the overall power delivered to the amplifiers. The combined signal is amplified using a \emph{Mini-Circuits ZX60-100VH+} power amplifier.

Coupling between the resonators is achieved purely inductively by physical proximity. Changing the coupling constants therefore involves adjusting the position of the resonators relative to each other. In the degenerate coupling case, we have a distance of roughly \SI{247}{\milli\metre} between the centers of the coils.

\section{Characterization and Tuning}

By detuning the frequencies of two resonators from that of the third, we can characterize the individual properties of that third resonator. Fitting the response curve to a frequency sweep across resonance for $U_\mathrm{d} = \SI{1}{\milli\volt}$, we extract the resonance frequency $\omega_0$ and the damping rate $\Gamma$ of the resonator in the linear regime. A fit to the amplitude response of a parametric frequency sweep (from low to high frequencies) yields the nonlinearity constant $\beta$ as well as the parametric threshold $U_\mathrm{th}$.

We find that due to three-wave mixing, the oscillation at $\omega$ drives a small response at $2\omega$, thereby reducing the amplitude at $\omega$ below the value expected from standard treatments~\cite{lifshitz2008nonlinear,Eichler_Zilberberg_book}. To compensate for this effect, we combine the amplitude response $A_\mathrm{r}$ at $\omega$ and $A_\mathrm{l}$ at $2\omega$ to $A_\mathrm{c} = \sqrt{A_\mathrm{r}^2 + A_\mathrm{l}^2}$ before fitting. The result agrees well with the expected amplitude response. We therefore use $A_\mathrm{c}$ for all parameter extraction from frequency-dependent data. However, amplitudes quoted from stochastic experiments at a fixed frequency (for Fig.~4 and Fig.~\ref{fig:figS1}) are ``raw values'' $A_\mathrm{r}$.

Tuning is required to make the resonator properties approximately identical. First, the resonance frequency, damping, and parametric threshold are measured for each resonator individually as a function of the bias voltage. Then, we keep the bias voltage $U_\mathrm{b}$ of one resonator fixed at $\SI{5}{\volt}$ and iteratively adjust the drive loops, damping, and $U_\mathrm{b}$ of the other resonators until all parameters match as closely as possible. The resulting tuning is good enough for our requirements with an error below \SI{0.01}{\percent} for the resonance frequency (best matching parameter) and roughly \SI{10}{\percent} for $U_\mathrm{th}$ (worst matching parameter).

Pairwise coupling constants are extracted by comparing the individual resonance frequency $\omega_i$ of one resonator (extracted by detuning the other two) to the symmetric normal mode frequency $\omega_\mathrm{s}$ of the pair it forms with a second resonator, while the third resonator remains far detuned. The symmetric mode frequency is found by fitting to the response of a frequency sweep while driving both resonators in phase. From the two measured frequencies, the coupling constant follows approximately as $J = \omega_i^2 - \omega_\mathrm{s}^2$ for $\omega_i\gg \omega_i-\omega_\mathrm{s}$. For the case of non-degenerate coupling, we obtain $J_{12} = \SI{-1709(63)e9}{\hertz\squared}$, $J_{23} = \SI{-1108(57)e9}{\hertz\squared}$, and $J_{31} = \SI{-1473(123)e9}{\hertz\squared}$. In the case of degenerate coupling, we extract the coupling constant from the frequency splitting observed in the parametric response, see Fig.~2(b) in the main text. We obtain $J = (\omega_\mathrm{a}^2 - \omega_\mathrm{s}^2) / 3 = \SI{-1107(4)e9}{\hertz\squared}$, with $\omega_\mathrm{a}$ and $\omega_\mathrm{s}$ the anti-symmetric and symmetric mode frequencies, respectively.

\section{Evaluation of switching data}

Our stochastic sampling method relies on activated switching between phase states, which can be triggered by applying uncorrelated white noise signals to each of the resonators in addition to the parametric drive. All switching measurements were taken with a parametric drive of $U_\mathrm{p} = \SI{400}{\milli\volt}$ at a frequency $\omega/\pi = \SI{3212960}{\hertz}$ and with a noise standard deviation of $\sigma = \SI{137.88}{\milli\volt}$. This noise intensity was selected because it leads to switching rates on the order of \SI{80}{\hertz}, which can be conveniently measured with our lock-in amplifier. Between the switches, we measure a standard deviation $\sigma_u \approx \SI{1}{\milli\volt}$ of the $u_i$-quadrature responses shown in Fig.~4(a) of the main text.

\begin{figure*}[t]
	\includegraphics[width=\textwidth]{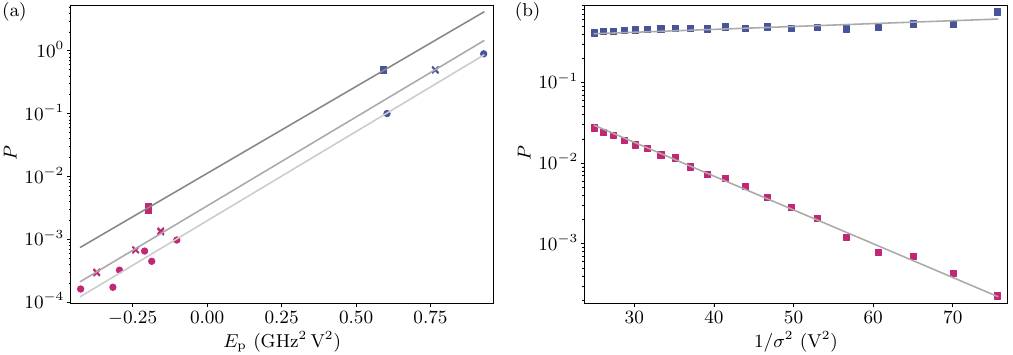}
	\caption{Exponential dependence of the occupation probabilities $P$ on energy and noise. (a)~Occupation probabilities $P$ as function of energy $E_\mathrm{p}$ for all three cases (i-iii) shown in Fig.~4 of the main text: degenerate coupling results (case i) are shown as squares, different $J_{ij}$ (case ii) as x, and data with additional external force (case iii) as dots. Gray lines are fits of $P(E_p) = c_1e^{E_p/c_2}$ with fit parameters $c_{1,2}$. For cases i we find $c_1 = \num{1.12(3)e-2}$ and $c_2 = \SI{0.157(2)}{\giga\hertz\squared\volt\squared}$, for case ii: $c_1 = \num{3.42(7)e-3}$ and $c_2 = \SI{0.153(1)}{\giga\hertz\squared\volt\squared}$ and for case iii: $c_1 = \num{1.99(2)e-3}$ and $c_2 = \SI{0.153(4)}{\giga\hertz\squared\volt\squared}$. At low energies, we observe deviations of the data with external force (dots) from the fit, indicating a potential limitation of the the Ising equivalence, see~\ref{sm: ising derivation}. Symmetric states $(\up\up\up)$ and $(\down\down\down)$ are shown in blue, all other states in pink. (b)~Measurement of the occupation probabilities as function of applied noise standard deviation $\sigma$ with degenerate coupling of $J \approx \SI{1330e9}{\hertz\squared}$ at $U_p = \SI{416}{\milli\volt}$ and $\omega/\pi = \SI{3211373}{\hertz}$. Here, we only plot the states $(\down\down\down)$ in blue and $(\down\down\up)$ in pink. Gray lines are a fit of $P(\sigma, E_p) = c_1'e^{(E_p+c_2')c_3'/\sigma^2}$ with fit parameters $c_{1,2,3}'$ found as $c_{1}' = \num{0.33(1)}$, $c_2' = \SI{-0.660(9)}{\giga\hertz\squared\volt\squared}$ and $c_3' = \SI{10.64(6)}{\giga\per\hertz\squared}$.}
	\label{fig:figS4}
\end{figure*}

For the stochastic sampling data, all three resonators' responses are recorded for $t_\mathrm{m} = \SI{400}{\second}$ with a sampling rate of $df = \SI{20.081}{\kilo\hertz}$ for each dataset. Using a thresholding algorithm with two circular thresholds as described in Ref.~\cite{margiani2021fluctuating}, the data is translated into spin states (up or down) for each resonator individually and then combined, yielding the system's state configuration for each point in time. The total dwell time $\tau$ spent in a specific configuration can then be calculated as
$\tau = n_\mathrm{s} / df$, with $n_\mathrm{s}$ the number of samples assigned to that configuration over the whole measurement. The resulting times are converted to the occupation probability $P = \frac{\tau}{t_m}$ and compared with the energies of an Ising Hamiltonian. The model uses the coupling constants $W_{ij}$, derived from the experiment as $W_{ij} = \frac{J_{ij} A^2}{2}$, with $A=A_\mathrm{Max}$ the largest stationary oscillation amplitude measured in the dataset. This scaling ensures the same units between the Ising model energy $E_\mathrm{I}$ and the rotating-frame quasi-energy $E_\mathrm{p}$, see Section~\ref{sm: ising derivation}. To verify that our system actually follows Boltzmann statistics, we fit an exponential model in the energies $E_\mathrm{p}$ to the resulting occupation probabilities $P$ in Fig.~\ref{fig:figS4}.

\section{Different estimators for the Ising state order}

In a Boltzmann machine, the state with the highest $P$ is expected to have the lowest energy. This can be used to solve the Ising problem. In this work, we identify the state with the lowest Ising energy $E_\mathrm{I}$ by ordering the states according to their $P$. In Ref.~\cite{Goto_2018}, the highest $P$ corresponds to the lowest $E_\mathrm{I}$, as in the Ising model. In our work, we show that depending on the relative sign of $\beta$ and $J$, the Ising solution can also correspond to the lowest $P$.

\begin{figure*}[t]
	\includegraphics[width=\textwidth]{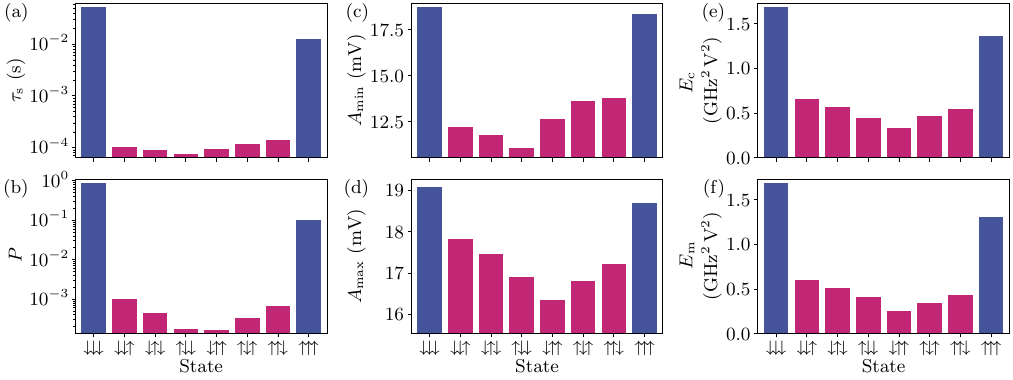}
	\caption{Estimators for the Ising ground state in the presence of an external force simulating a magnetic field in the Ising Hamiltonian. (a)~Average lifetime of the system in each state configuration. (b)~Occupation probably of each state configuration. (c)~Lowest of the three oscillation amplitudes in a state configuration. (d)~Largest of the three oscillation amplitudes in a state configuration. (e)~Full rotating-frame quasi-energies assuming equal amplitudes over all states and resonators. (f)~Full rotating-frame quasi-energies calculated using measured amplitudes.}
	\label{fig:figS1}
\end{figure*}

Besides the occupation probability $P$, several other measurables can be used as estimators for the ordering of the Ising energies. Possible examples are the average lifetime $\tau_\mathrm{s}$ of a state~\cite{margiani2021fluctuating} or the response amplitudes of the resonators in the different states. We found that all these estimators lead to the same ordering in our system as long as the symmetry between the phase states is not broken, that is, without an external force. However, when applying an external force to simulate an external magnetic field, some of the estimators result in a different ordering than the Ising Hamiltonian, cf. different results in Fig.~\ref{fig:figS1}. For example, the lifetime $\tau_\mathrm{s}$ produces a different hierarchy between the states that does not correctly predict that of the Ising Hamiltonian. This is likely due to the fact that $\tau_\mathrm{s}$ depends only on the probability per unit time of \textit{leaving} a certain state. This probability can be low for a state that is close in quasienergy to a saddle node, allowing for frequent activation out of the state. Therefore, we find that $\tau_\mathrm{s}$ is not a good estimator for $E_\mathrm{I}$, in contrast to $P$.

Of particular interest is the role of the oscillation amplitudes in different state configurations. In particularly simple cases, we found that these amplitudes can already be used to estimate the Ising energy order~\cite{margiani2021fluctuating}. However, for the more involved examples presented here, this is generally not true. In Sec.~\ref{sm: ising derivation}, we will show how the amplitudes are expected to influence the rotating-frame quasi-energies. For the comparison in Fig.~4 of the main text, we assumed for simplicity that all amplitudes are identical. In Fig.~\ref{fig:figS1}(e) and (f), we compare the quasi-energies with equal amplitudes ($E_\mathrm{c}$) versus those with measured, unequal amplitudes ($E_\mathrm{m}$). For equal amplitudes, the state energies in $E_\mathrm{c}$ differ from those in $E_\mathrm{p}$ only by a constant summand. see the two first right-hand side terms in Eq. (2) of the main text. We find that unequal amplitudes can modify the quasi-energy differences, potentially making the mapping between the KPO network and an Ising Hamiltonian invalid. Experimentally, this implies that the simulator should be employed in a regime far beyond the parametric threshold, where the amplitudes become approximately identical. Understanding exactly how the different factors impact the measured result and in which situations various metrics can be used as an indicator for the Ising ground state, remains a challenge for future work.

\begin{figure*}[t]
	\includegraphics[width=0.5\textwidth]{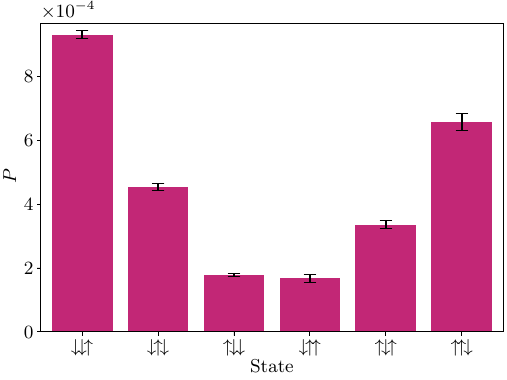}
	\caption{Occupation probability $P$ for the six `middle' states in Fig.~4(b), case iii, with error estimation. Symmetric configurations are omitted for better visibility of the small values. Errors where calculated from the standard deviation of $\tau$ between 16 separate parts of the data set with length $t_\mathrm{p} = \frac{t_m}{16}$ each. The standard deviation calculated from the 16 individual subsets were divided by 4 to estimate the standard deviation of the full set.}
	\label{fig:figS2}
\end{figure*}

In Fig.~\ref{fig:figS2}, we display the values of $P$ for the six middle states from Fig.~4(b), case iii. To make small differences between these states with low $P$ more visible, we omit here the symmetric states $(\up \up \up)$ and $(\down \down \down)$. In order to gain a heuristic estimation of the uncertainties for these values, we divide the full data set into 16 subsets with equal lengths $t_\mathrm{p} = \frac{t_m}{16}$. This allows us to calculate a standard deviation for $\tau$ of each individual state, indicating how significant the measured differences between the states are. The standard deviations for the full data set are then estimated as 1/4 of the standard deviations of each subset (as $4=\sqrt{16}$). From the figure, we can see that the differences between all states are much larger than the corresponding standard deviations, except for the two lowest and nearly degenerate states.

%The amplitudes of the individual resonators take a special role in our derivation of the Ising model Hamiltonian from the rotating-frame Hamiltonian of the parametric resonator network (see Sec.~\ref{sm: ising derivation}). When not entirely degenerate, they introduce a deviation from the Ising model that could potentially be significant. To validate that assuming equal amplitudes in the calculations is permissible for our experiment, 

\section{Network of parametric oscillators as an Ising machine}
\label{sm: ising derivation}

The network of coupled parametric oscillators can be described by the following Hamiltonian:
\begin{equation}\label{SM: Classical Hamiltonian}
  H = \sum_i \left[\frac{p_i^2}{2m_i} + \frac{m_i\omega_i^2}{2}(1-\lambda\cos(2\omega t))x_i^2 + \frac{\beta_i}{4}x_i^4-F_i\cos(\omega_i t)x_i\right] - \sum_{j>i} J_{ij}x_ix_j,
\end{equation}
where $x_i$ and $p_i$ represent the position and momentum of the $i$-th oscillator, respectively. Each oscillator has mass $m_i$ and natural frequency $\omega_i$. The system is driven parametrically at frequency $2\omega$ with strength $\lambda$ in addition to an external forcing at frequency $\omega$ with amplitude $F_i$, and it exhibits a Duffing nonlinearity with coefficient $\beta_i$. The oscillators are coupled through terms with strength $J_{ij}$. We consider unitless masses $m_i=1$ in our analysis. This Hamiltonian leads to the coupled equations of motion shown in Eq.~(1) of the main text, where we additionally included damping terms with rate $\Gamma_i$ to account for energy dissipation in the experimental system. We assume identical resonators, with $\beta_i=\beta$ and $\omega_i=\omega_0$.

As discussed, the primary response of the resonators is at frequency $\omega$, corresponding to half the parametric drive frequency. Hence, it is convenient to express the dynamics in terms of quadrature variables at $\omega$:
\begin{align} \label{sm: van_der_pol}
  x & = u \cos(\omega t) + v \sin(\omega t),                       \\
  p & = - \omega \, u \sin(\omega t) + \omega \, v \cos(\omega t),
\end{align}
where $u$ and $v$ represent the quadrature components of the oscillation. The transformation preserves the canonical structure with Poisson bracket $\{u, v\} = \frac{1}{\omega}$ following from $\{x, p\} = 1$. This canonical transformation can be derived from the type-II generating function
\begin{equation}
  S_2 \left(x , v , t\right) = \omega v x \sec \left(\omega t\right) - \frac{\omega}{2} \left(v^2 + x^2\right) \tan \left(\omega t\right).
\end{equation}
Indeed, using the relations $u=\frac{1}{\omega} \pdv{S_2}{v}$ and $p=\pdv{S_2}{x}$, we obtain Eq.~\eqref{sm: van_der_pol}. The factor $1/\omega$ in $u=\frac{1}{\omega} \pdv{S_2}{v}$ ensures consistent units, with both $u$ and $v$ having dimensions of length (resp. voltage in the model we use in the main text).

The Hamiltonian for a single resonator from Eq.~\eqref{SM: Classical Hamiltonian} can be expressed in the new basis variables as
\begin{align} \label{sm: cl rotating Hamiltonian}
  \tilde{H}(u,v,t) & = H(x,p,t) + \pdv{S_2}{t}\nonumber
  \\
  & = \frac{3}{32} \beta (u^2+v^2)^2 + \frac{\lambda\omega_0^2}{8}(v^2-u^2) - \frac{1}{4}(u^2+v^2) (\omega^2- \omega_0^2)
  \nonumber\\
  & \quad + \frac{1}{4} v \sin(2 \omega t) \left[\beta (u^2 + v^2)u + 2 (\omega^2 -  \omega_0^2)u-2F\right]
  \nonumber\\
  & \quad + \frac{1}{8} \cos(2 \omega t) \left[\beta( u^4- v^4 ) - 2(\lambda\omega_0^2 + \omega^2 - \omega_0^2)(u^2+v^2)-4Fu\right]
  \nonumber\\
  & \quad + \frac{1}{8} u v \sin(4 \omega t) \left[\beta (u^2 - v^2) - 2 \lambda\omega_0^2\right]
  \nonumber\\
  & \quad + \frac{1}{32} \cos(4 \omega t) \left[4 \lambda\omega_0^2 (v^2 - u^2) + \beta (u^4 - 6 u^2 v^2 + v^4)\right]\,.
\end{align}
This transformation splits the dynamics into two parts: rapidly counter-rotating terms oscillating at frequencies $2\omega$ and $4\omega$, and slower stroboscopic dynamics co-rotating with frequency $\omega$. Assuming we are only interested in the stationary dynamics of the parametric resonator network, we can derive an effective Hamiltonian $H_{\mathrm{eff}}$ by averaging $\tilde{H}$ over one stroboscopic period $T=\frac{2\pi}{\omega}$,
\begin{equation} \label{sm: hamiltonian KB}
  H_{\mathrm{eff}} =\frac{1}{T}\int_0^{T} \tilde{H} \dd{t},
\end{equation}
allowing us to study the system's stroboscopic evolution. This procedure can be formalized into an expansion where Eq.~\eqref{sm: hamiltonian KB} serves as a first-order approximation \cite{burshteinHamiltonian1962,Seibold_2025}. The same expansion can be derived for the full equations of motion [cf. Eq.~(1) in the main text], including dissipation using a near-identity transformation~\cite{holmesSecond1981,Seibold_2025}.
To first order, we obtain the effective Hamiltonian
\begin{multline} \label{sm: hamiltonian KB first order}
	H_\mathrm{eff}(u,v) = \sum_i \frac{3}{32} \beta (u_i^2+v_i^2)^2 + \frac{\lambda\omega_0^2}{8}(v_i^2-u_i^2) \\
	- \frac{1}{4}(u_i^2+v_i^2) (\omega^2- \omega_0^2) -\frac{F_i}{2}u_i- \sum_{j \neq i} \frac{J_{ij}}{2} (u_i u_j + v_j v_i)\,.
\end{multline}

To demonstrate that a network of parametric oscillators maps to an Ising model, we first express the effective Hamiltonian in polar coordinates. We introduce amplitudes $A_i$ and phases $\phi_i$ variables through $u_i = A_i \cos(\phi_i)$ and $v_i = A_i \sin(\phi_i)$, where $A_i>0$ and $0 \leq \phi_i \leq 2\pi$. Using a canonical transformation with type-I generating function $S_1(u_i, \phi_i) = \frac{1}{2} u_i^2 \cot(\phi_i)$, the decoupled part of the Hamiltonian at a particular state $i$ becomes:
\begin{equation}
  H_0 =\sum_i \frac{3}{32} \beta A_i^4 - \frac{1}{8} \left( 2 \omega^2 - 2 \omega_0^2 + \lambda\omega_0^2 \cos (2 \phi_i) \right) A_i^2 - \frac{F_iA_i}{2}\cos(\phi_i).
\end{equation}
The coupling-dependent terms transform to:
\begin{equation}
  H_J = -\sum_{j \neq i} \frac{J_{ij}}{2} A_i A_j \left( \cos(\phi_i) \cos(\phi_j) + \sin(\phi_i) \sin(\phi_j) \right),
\end{equation}
such that $H_\mathrm{eff}= H_0 +H_J$. In the absence of dissipation, the parametric phase states are constrained to $v_i = A_i \sin(\phi_i) = 0$, implying $\phi_i = 0$ or $\pi$. By defining classical spin variables $\sigma_i = \cos(\phi_i)$, part of the Hamiltonian resembles that of an Ising model:
\begin{equation} \label{sm: Ising Hamiltonian}
  H_\mathrm{p} = -\sum_{j \neq i} \frac{J_{ij}}{2} A_i A_j \sigma_i \sigma_j -\sum_i \frac{F_i}{2}A_i\sigma_i.
\end{equation}
The energies $E_\mathrm{p}$ shown in the main text correspond to the values of $H_\mathrm{p}$ for particular states.
When the oscillator amplitudes are equal, this exactly maps to the Ising Hamiltonian up to an energy rescaling with coefficients $W_\mathrm{ij}=\frac{J_{ij} A^2}{2}$ and $B_\mathrm{I}=\frac{F}{2A}$ where $A\approx A_i$ [see Eq. (5) of the main text]. Using perturbation theory in the weak coupling limit $J_{ij}=J \ll 1$, the amplitudes are approximately equal:
\begin{equation}
  A_i \approx\sqrt{\frac{2\Lambda}{3\abs{\beta}}} - \sqrt{\frac{2 }{3\abs{\beta}\Lambda} } J+ \mathcal{O}(J^2)\,.
\end{equation}
with $\Lambda=(2 \omega^2 - 2 \omega_0^2) + \mathrm{sign}(\beta)\lambda\omega_0^2$. Therefore, when either $ \Lambda \gg 1$ or $J_{ij} \ll 1$, the phase states of our oscillator network effectively behave as classical spins in an Ising model.

Equation~\eqref{sm: Ising Hamiltonian} shows that the mapping between a KPO network and an Ising Hamiltonian works best when all amplitudes $A_i$ are identical. This ideal situation is approximated by driving the network far beyond the parametric threshold, where the relative impact of $W_{ij}$ becomes small.

\section{Relative signs of $\beta$ and $J$}
\begin{figure*}[t]
	\includegraphics[width=\textwidth]{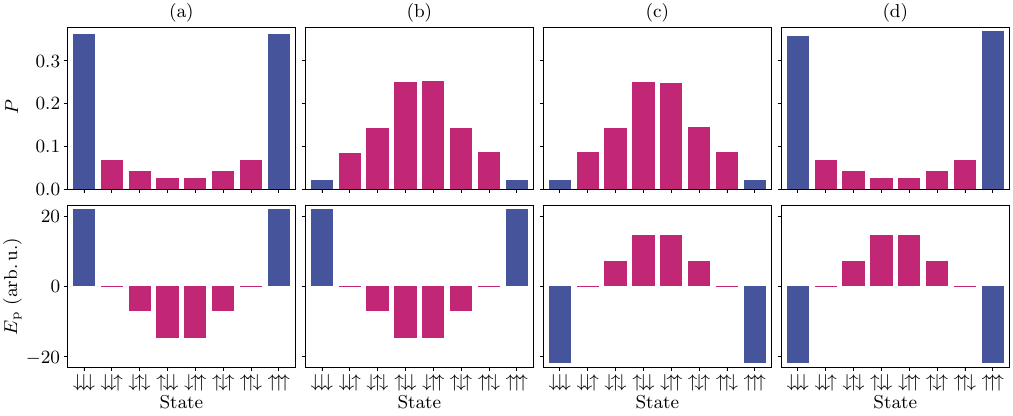}
	\caption{Effect of the sign of $\beta$ and $J$ on the occupation probabilities $P$ (top row) and rotating frame quasi-energies $E_\textrm{p}$ (bottom row). Plots show schematics for (a) negative $\beta$ and $J$, (b) positive $\beta$ and negative $J$, (c) negative $\beta$ and positive $J$, and (d) positive $\beta$ and $J$.}
	\label{fig:figS3}
\end{figure*}

In Fig.~\ref{fig:figS3}, we schematically show case ii from Fig.~4(b) for positive and negative $\beta$, as well as positive and negative $J$. The comparison between $P$ and $E_\mathrm{p}$ illustrates how the relative sign between $\beta$ and $J$ influences whether the states with highest or lowest energy are most occupied.

To reach a better intuition about the effects of the sign of $\beta$ and $J$, we propose to form a new effective Hamiltonian $\tilde{H}_\mathrm{eff}=H_\mathrm{eff}\times \mathrm{sign}(\beta)$. When considering the full Hamiltonian in Eq. (S7), we can make the following statements about the corresponding therms in $\tilde{H}_\mathrm{eff}$: the first term on the right-hand side, corresponding to the nonlinear potential contribution, remains always positive. The second term, giving the quadratic potential term, always has a negative contribution in one of the quadratures. Hence, in $\tilde{H}_\mathrm{eff}$ the axis on which the stable solutions lie is rotated by \SI{90}{\degree} when $\beta<0$. The third term contains the detuning and is inverted for $\beta<0$. For small detuning, this merely produces a global quasi-energy shift and can be neglected when considering energy differences. Finally, the coupling term is inverted with the sign of $\beta$. The transformation effectively removes the possibility of maxima as solutions, making every solution a minimum. This allows us to apply the intuition from equilibrium physics despite looking at a rotating frame quasi-Hamiltonian of a driven system.

\section{Identifying the Ising regime}

In Ref.~\cite{ameye2025parametricinstabilitylandscapecoupled}, analytical expressions for bifurcation lines that characterize the stability of coupled Kerr parametric oscillator (KPO) networks were analyzed. These bifurcation lines define the boundaries of the region where the system exhibits a solution space consisting of \(2^N\) stable Ising states.

For the specific case of \(N=3\) coupled KPOs, as studied experimentally in the main text of this work, there will be two bifurcation lines that bound the Ising region in the $\lambda-\omega$ stability plot:
\begin{equation}
  \lambda_\mathrm{l} = 2\sqrt{\gamma^2\omega^2+(\omega_0^2-\omega ^2)^2}/\omega_0^2 \qq{and}   \lambda_\mathrm{r} = 2\sqrt{\gamma^2\omega^2+\left(\omega_0^2 +\left(1+\Omega\frac{\operatorname{sign}(\beta)}{2}\right) J-\omega ^2\right)^2}/\omega_0^2,
\end{equation}
with $\Omega=2 \sqrt{6 \sqrt{3}+9}$. The boundary of the Ising regime is therefore:
\begin{equation} \label{sm: Ising boundary}
  \lambda_\mathrm{Ising}=
  \begin{cases}
        \lambda_\mathrm{l} \qq{when} \omega < \frac{\sqrt{4\omega_0^2+(\Omega-2) J}}{2} \\
    \lambda_\mathrm{r}   \qq{when}  \omega >  \frac{\sqrt{4\omega_0^2+(\Omega-2) J}}{2}  \
  \end{cases}
\end{equation}
This coincides with the experimentally observed and numerically computed boundary of the Ising regime in Fig.~(2) of the main text. When the system is within the region defined by Eq.~\eqref{sm: Ising boundary}, it will exhibit a solution space consisting of \(2^N\) stable parametric states.

For N equally coupled KPO's the bifurcation line has the same form as Eq.~\eqref{sm: Ising boundary} and is instead shifted by  Ref.~\cite{ameye2025parametricinstabilitylandscapecoupled}:
\begin{equation}\label{sm: shift}
  \Omega(N,M)=\sqrt{3 \Xi+ \Sigma}+\sqrt{2  \Sigma-3 \Xi-\operatorname{sign}(\beta)\frac{2  N \left(N^2-9 M^2\right)}{\sqrt{3 \Xi+ \Sigma}}}  ,
\end{equation}
with $\Xi=\sqrt[3]{- \left(M^3-M N^2\right)^2}$, $\Sigma=\left(3 M^2+N^2\right)$ and $M=|n-m|$ the absolute difference between resonators with phase $\phi$ and $\phi+\pi$. It is assumed that $n> m$. Hence, the expression predict $\left\lfloor \frac{N+2}{2} \right\rfloor$ bifurcation lines for a system with N equally coupled KPO's.

\bibliography{aipsamp}% Produces the bibliography via BibTeX.